\documentclass[%
 aip,
 jmp,%
 amsmath,amssymb,
 preprint,%
]{revtex4-2}

\usepackage{graphicx}
\usepackage{dcolumn}
\usepackage{bm}

\usepackage[utf8]{inputenc}
\usepackage[T1]{fontenc}
\usepackage{mathptmx}
\usepackage{etoolbox}

\usepackage{siunitx}
\usepackage{lipsum}
\DeclareSIUnit\torr{Torr}
\DeclareSIUnit\oersted{Oe}

\begin{document}


\title[Synchronization of propagating spin waves in spin Hall oscillators: A micromagnetic study]{Synchronization of propagating spin waves in spin Hall oscillators: A micromagnetic study}


\author{Mohammad Haidar}
\affiliation{Department of Physics, American University of Beirut, Riad El-Solh,, Beirut, 1107-2020, Lebanon}%
 \email{mh280@aub.edu.lb}

\date{\today}

\begin{abstract}
In this study, we investigate the synchronization of propagating spin waves in a novel spin torque oscillator device layout using micromagnetic simulations. This design enables individual probing of the dc current in each oscillator, allowing precise control over the synchronization state and providing direct phase measurement. Our findings reveal that two adjacent oscillators achieve phase locking when they maintain a constant phase difference, either in phase or anti-phase, depending on their separation distance. This work offers new insights into STO synchronization mechanisms and paves the way for improved control and functionality in spintronic devices.
\end{abstract}

\maketitle

\section{Introduction}

The synchronization of oscillatory systems is a fundamental phenomenon observed in nature, where mutual interactions among oscillators lead to a collective rhythm \cite{Acebron2005}. In the field of spintronics, spin torque oscillators (STOs) have emerged as a versatile platform for studying nonlinear dynamics and exploring practical applications, such as microwave signal generation \cite{slavin2009,DemidovSTO2010, Slavin2005,Houshang2018,Haidar2019}, neuromorphic computing \cite{Torrejon2017, Tsunegi2019, Romera2018}, and wireless communication \cite{Dieny2020}. Synchronization among spin-torque oscillators is a particularly intriguing phenomenon, as it enhances their output power, spectral purity, and coherence \cite{Mancoff2005, Pufall2006, Slavin2006,Berkov2013}. Various coupling mechanisms, including magnetic dipolar interactions \cite{Erokhin2014,Castro2022,Martins2023}, spin waves \cite{Madami2011,Kendziorczyk2014, Houshang2016,Kendziorczyk2016}, and electrical coupling \cite{Lebrun2017,Taniguchi_2018}, have been explored to achieve both short- and long-range synchronization. Additionally, advances in the long-range synchronization of spin-torque oscillators using spin waves have opened doors for creating highly interconnected oscillator networks with applications in complex problem-solving, such as Ising machines \cite{Houshang2022, Litvinenko2023}.  
Among different layout of oscillators, the spin Hall nano-oscillators (SHNOs), a new class of oscillators consisting of a ferromagnet/heavy metal (FM/HM) bilayer, are promising candidates for synchronization in one- and two-dimensional arrays \cite{Awad2016, Zahedinejad2020}. However, in these devices, synchronization is dictated by dipolar fields. With the recent development of SHNOs and including perpendicular magnetic anisotropy (PMA) materials as the ferromagnetic layer, propagating spin waves can be excited \cite{Fulara2019, Succar2023}, offering potential for long-range synchronization \cite{kumar2025}. Moreover, recent advancements have highlighted the potential of SHNOs arrays as tunable nanoscale signal sources, capable of synchronized producing coherent spin-wave interference patterns \cite{haidar2024}. The long-range synchronization of SHNOs arrays via spin wave coupling faces significant challenges due to the rapid decay of the amplitude of the propagating spin wave packet in metallic ferromagnets, which is less than 1 $\mu m$. Moreover, in spin Hall devices, the effective damping of the ferromagnetic/heavy metal bilayers increases tremendously due to strong spin-orbit coupling \cite{Ranjbar2014,Yuli2015, Haidar2021}; hence, the propagation length shrinks significantly, resulting in an effective synchronization over a distance below \SI{500}{nm}. Controlling the relative phase between the oscillators represents another fundamental challenge for understanding synchronization in nanoscale devices \cite{Slavin20092, Ruotolo2009}, where significant progress in phase-resolved techniques has been made to improve accessibility at the nanoscale.

In this work, we study the synchronization of propagating spin waves in a novel spin Hall nano-oscillators (SHNOs) device using micromagnetic simulations. This layout offers two key advantages: (i) the ability to probe the dc current of each oscillator individually, enabling control over the synchronization state, and (ii) direct measurement of the phase of each oscillator. Our results demonstrate that phase locking between two adjacent oscillators occurs when they maintain a constant phase difference, either in-phase or anti-phase, depending on their separation distance.
\begin{figure*}[t]
\begin{center}
\includegraphics[width=0.8\textwidth]{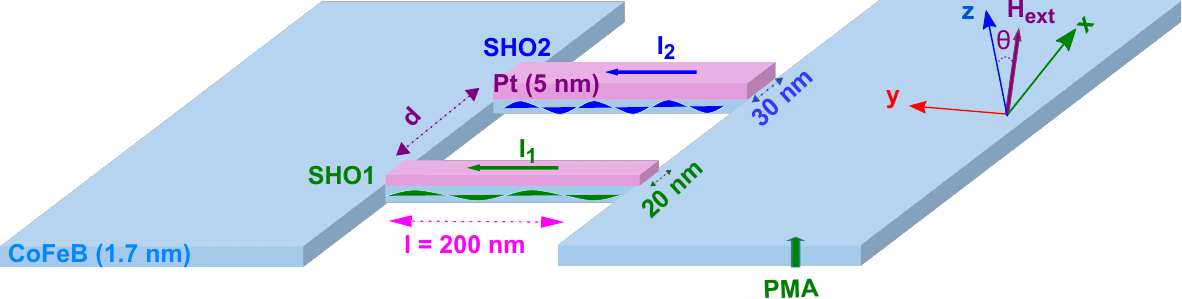}
\caption{(Color online) Schematic representation of two nanowire-based Spin Hall oscillators (SHO1 and SHO2). The oscillators consist of a CoFeB (\SI{1.7}{nm}) and Pt (\SI{5}{nm}) bilayer. SHO1 and SHO2 have widths (\textit{w}) of \SI{20}{nm} and \SI{30}{nm}, respectively, a length \text{l} = \SI{200}{nm}, and a separation distance \textit{d}. The schematic also illustrates the direction of the applied external magnetic field and the dc currents $I_{1}$ and $I_{2}$.}
\end{center}
\end{figure*}

\section{Micromagnetic simulations}
To initiate micromagnetic simulations, we model a layer comprising \SI{1.7}{nm} CoFeB patterned rectangular-shaped nanowires of length (\textit{l}) = \SI{200}{nm}. The two nanowires oscillators SHO1 and SHO2 have different widths of \textit{w} = \SI{20}{} and \SI{30}{nm},  respectively, separated by a distance (\textit{d}) between \SI{100}{} and \SI{600}{nm} as shown in Fig. 1. A \SI{5}{nm} Pt layer is deposited just over the nanowire with separate connection pads, allowing for an individual control of the current across the two nanowires. In-plane currents of $I_{1}$ and $I_{2}$ are applied along the y-axis in the \SI{20}{nm} and \SI{30}{nm} oscillators, respectively, where a high current density is concentrated within the nanowire region. Micromagnetic simulations were done using the mumax$^{3}$ solver. In these simulations, we adopt a rectangular mesh with dimensions of \SI{2000}{}$\times$\SI{2000}{}$\times$\SI{1.7}{nm^3} and a cell size of \SI{3.9}{}$\times$\SI{3.9}{}$\times$\SI{1.7}{nm^3}. We calculate the electrical current density and the corresponding Oersted field in each oscillator. Then, we convert the electrical current density (\text{J}${e}$) to the spin current density (\text{J}${s}$) via the relation $J_\text{s} = \theta_\text{SH} J_\text{e}$, where $\theta_\text{SH}$ represents the spin Hall angle of Pt and is equal to \SI{0.1}{}. Furthermore, in the simulation, we assume that the injected spin current predominantly induces a damping-like torque of the Slonczewski form. For micromagnetic simulations, the CoFeB/Pt nanowires are characterized by a saturation magnetization $\mu_{0}$$M_\text{s}$ of \SI{0.9}{T}, a perpendicular magnetic anisotropy $K_{u}=$ \SI{0.07}{}$MJ/m^3$, a gyromagnetic ratio $\gamma/2\pi$ of \SI{30}{GHz/T}, and an exchange stiffness \text{A} of \SI{15}{pJ/m}, consistent with experimental findings \cite{Zahedinejad2018apl}. A Gilbert damping coefficient $\alpha$ of \SI{0.022}{} is used in the CoFeB/Pt region, and \SI{0.01}{} outside.
The magnetization dynamics are simulated by integrating the Landau-Lifshits-Gilbert-Slonczewski (LLG-S) equation over \SI{150}{ns}. Excited mode frequencies and their spatial profiles are determined by performing a Fast Fourier Transform (FFT) of the time-domain data representing magnetization evolution.

\section{Results and discussion}
\begin{figure*}[t]
\begin{center}
\includegraphics[width=0.7\textwidth]{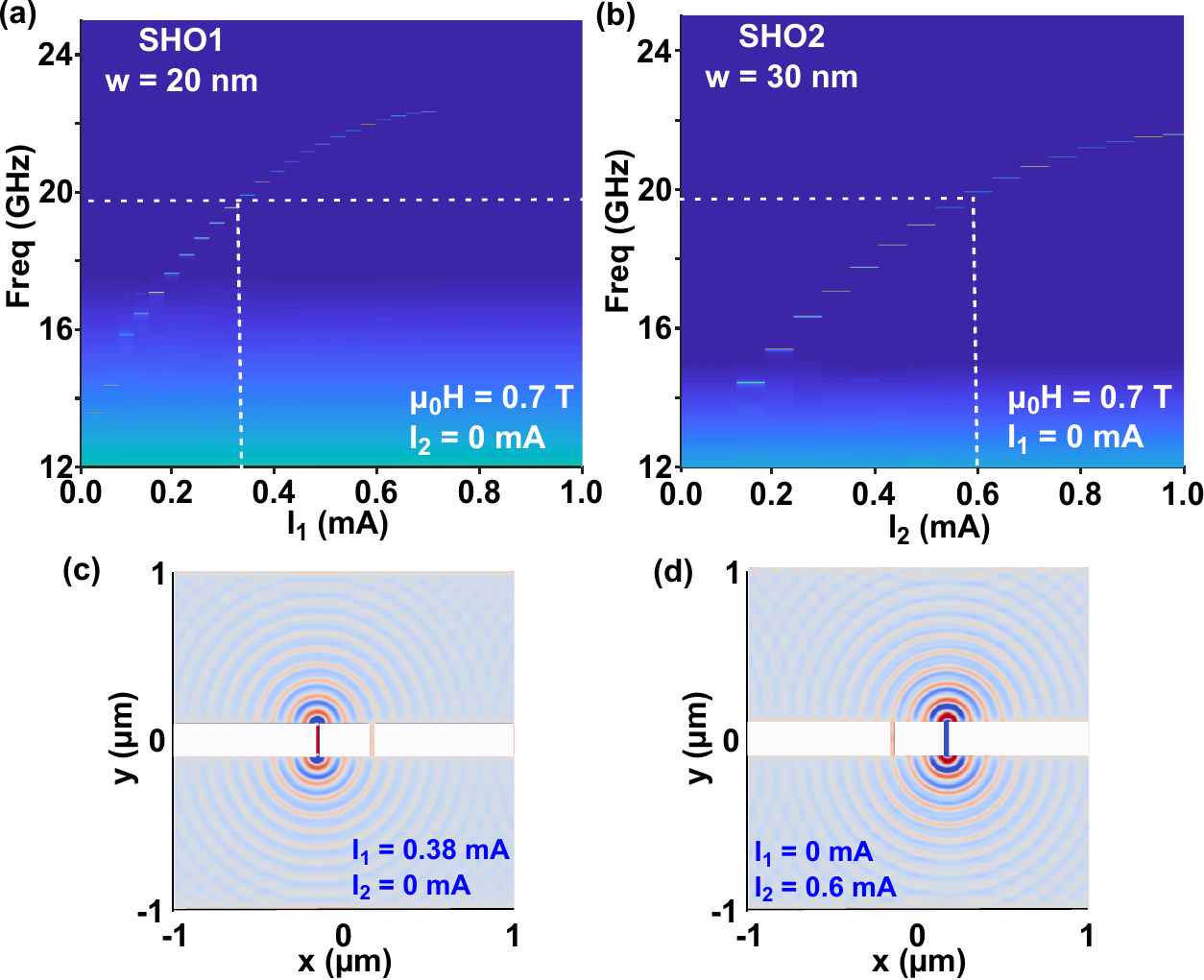}
\caption{(Color online) Power spectra color plot of spin-wave modes in (a) SHO1 and (b) SHO2 oscillators as a function of dc current $I_1$ and $I_2$. The white line highlighted the current $I_{1}$ and $I_{2}$ to generate a frequency of about \SI{19.9}{GHz}. The micromagnetic simulations were conducted under an applied magnetic field of \SI{0.7}{T} at an out-of-plane angle of \ang{75} with a perpendicular magnetic anisotropy  K$_{u} =$ \SI{0.07}{MJ/m^3}. Two-dimensional snapshots of the magnetization dynamics extracted at (c) ($I_1$ = \SI{0.38}{mA}, $I_2$ = \SI{0}{mA}) and (d) ($I_1$ = \SI{0}{mA}, $I_2$ = \SI{0.6}{mA}) corresponding to spin wave frequency of \SI{19.95}{GHz}. Note that the oscillation have identical wavelength of $\lambda$ = \SI{95}{nm}.}
\end{center}
\end{figure*}

The proposed layout has the advantage of controlling the injected dc current and reading the induced auto-oscillation characteristics using a spectrum analyzer or the individual phase using an oscilloscope connected to each oscillator experimentally. This device allows for additional tuning of oscillators between synchronized and unsynchronized states.
Before analyzing synchronization, we examine the auto-oscillations of each spin Hall oscillator (SHO1 and SHO2) individually. We perform micromagnetic simulations in which the devices are subjected to an external magnetic field applied at $75^{\circ}$ out-of-plane of magnitude $\mu_{0}H =$ \SI{0.7}{T}. The z-component of the magnetization (\textit{$m_z$}) is recorded over time while scanning the applied currents ($I_{1}$ in SHO1 or $I_{2}$ in SHO2) between \SI{0}{} and \SI{1}{mA}. Fig. 2(a) presents the power spectral density of auto-oscillations as a function of $I_{1}$, with $I_{2}$ set to \SI{0}{mA}. The onset current for auto-oscillation excitation is approximately \SI{0.15}{mA}, with oscillation frequencies ranging from \SI{14}{GHz} and \SI{23}{GHz}. Similarly, Fig. 2(b) shows the power spectral density for SHO2 as a function of $I_{2}$, with $I_{1}$ set to \SI{0}{mA}. A continuous auto-oscillation mode is observed, covering the same frequency range. The oscillatory modes in SHO1 and SHO2 correspond to the excitation of propagating spin waves driven by spin torque, exhibiting frequencies higher than the ferromagnetic resonance (FMR) frequency \cite{Fulara2019, Succar2023}. Due to the different widths of the two oscillators, the wavelengths of the excited spin waves differ; however, these wavelengths can be tuned by adjusting the dc current, as reported in \cite{Haidar2023}. For synchronization, both oscillators must oscillate at the same frequency or should have the same wavelength. The mode profiles of the auto-oscillations in SHO1 at $I_{1}=$ \SI{0.38}{mA} and in SHO2 at $I_{2}=$ \SI{0.6}{mA} are shown in Figs. 2(c) and 2(d). It is observed that propagating spin waves originate at the center of both SHO1 and SHO2 and extend over a considerable distance. Notably, at these current values, the propagating waves exhibit identical wavelengths of approximately \SI{95}{nm}. As a proof of concept, we will study the synchronization between the oscillators at a frequency of \SI{19.95}{GHz}, which is achieved at $I_{1}=$ \SI{0.38}{mA} and $I_{2}=$ \SI{0.6}{mA}.

\begin{figure*}[t]
\begin{center}
\includegraphics[width=0.9\textwidth]{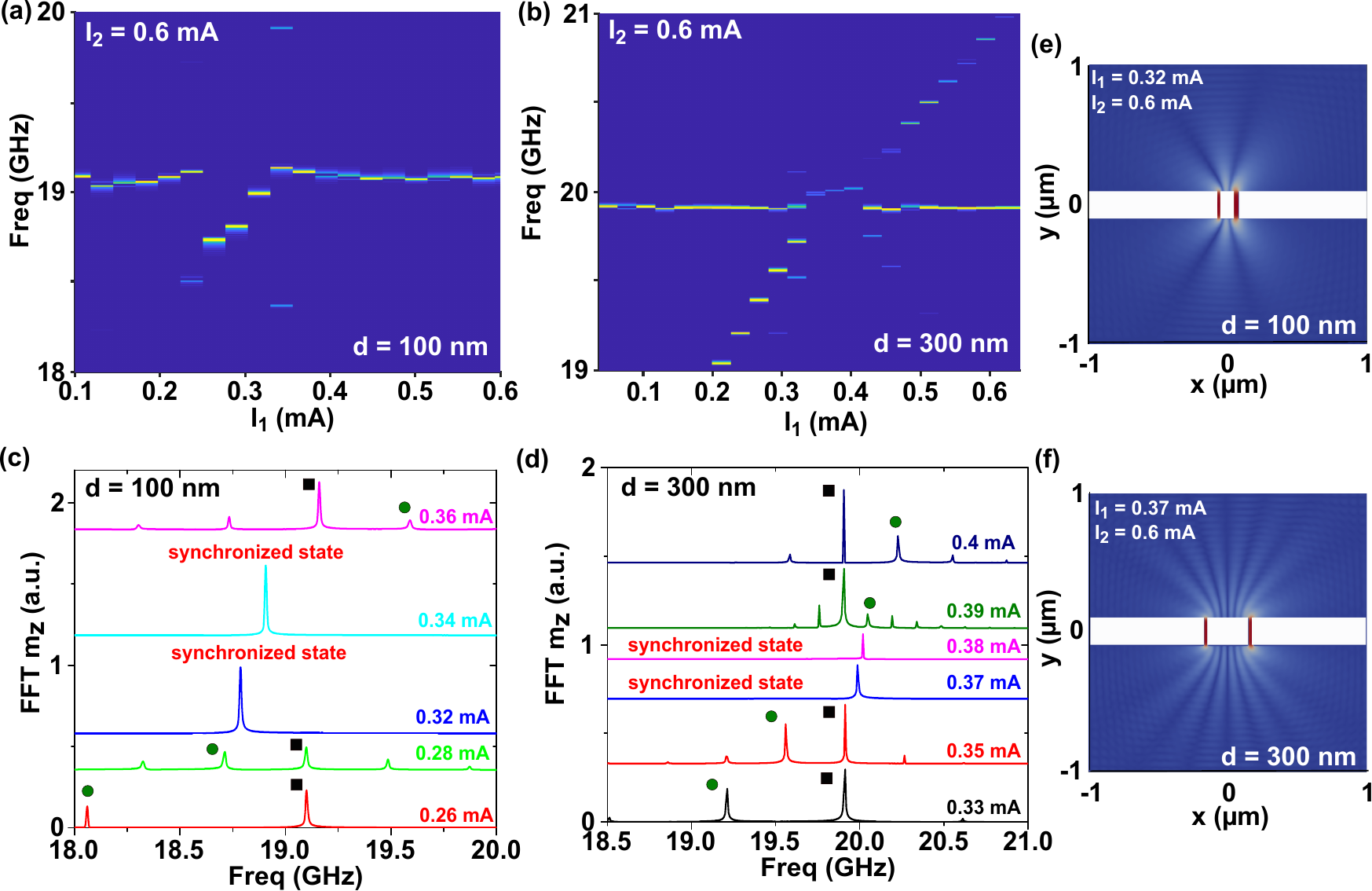}
\caption{(Color online)(a, b) Power spectral density color plot of spin-wave modes as a function of the dc current $I_{1}$, calculated at $I_{2}$ = \SI{0.6}{mA}, for a separation distance $d$ of (a) 100 nm and (b) 300 nm, demonstrating synchronization between SHO1 and SHO2. (c, d) FFT of $m_z$ as a function of frequency at different $I_{1}$ values. The closed square represents the oscillation of SHO2, while the closed circle represents the oscillation of SHO1 for (c) $d=$ \SI{100}{nm} and (d) $d=$ \SI{300}{nm}.
(e, f) Spin-wave intensity profiles showing interference patterns for (e) $I_{1}$ = \SI{0.3}{mA} and $I_{2}$ = \SI{0.6}{mA}, indicating in-phase synchronization, and (f) $I_{1}$ = \SI{0.36}{mA}and $I_{2}$ = \SI{0.6}{mA}, indicating anti-phase synchronization.}
\end{center}
\end{figure*}

To investigate the synchronization between the two oscillators, we fix the current $I_{2}$ in SHO2 at \SI{0.6}{mA}, corresponding to a frequency $f_{2}$. In the micromagnetic simulations, we vary the current $I_{1}$ in SHO1, leading to the emission of a propagating wave with a tunable frequency $f_{1}$. Figs. 3(a) and 3(b) present a zoomed-in view of the power spectral density as a function of $I_{1}$ for two devices with different separation distances between SHO1 and SHO2: 100 nm and 300 nm. The frequency $f_{2}$ remains constant, appearing as a continuous line, while $f_{1}$ exhibits a nonlinear blue shift with increasing $I_{1}$, consistent with the positive nonlinearity driving the propagating waves. At lower currents, the two STOs oscillate at different frequencies in both devices. As $I_{1}$ increases, $f_{1}$ gradually approaches $f_{2}$, where we observe a beating effect in $mz$ over time (not shown). A mutual synchronized state emerges at $I_{1}=$ \SI{0.3}{mA} for the 100 nm device [Fig. 3(a)] and at $I_{1}=$ \SI{0.36}{mA} for the 300 nm device [Fig. 3(b)]. To confirm synchronization, we analyze the FFT of $m_z$ as a function of frequency, where $f_{1}$ and $f_{2}$ are highlighted by circles and squares in Figs. 3(c) and 3(d) for d = \SI{100}{nm} and d = \SI{300}{nm}. The frequency spectrum reveals a single dominant frequency within the range $\Delta I_1 =$ \SI{0.2}{mA}, indicating that SHO1 and SHO2 are phase-locked and oscillate coherently at a common frequency, $f_{1} = f_{2} = f_{s}$. A key observation is that the nature of the synchronized state differs between the two devices, depending on the separation distance. For d = \SI{100}{nm}, the synchronized frequency is lower than $f_{2}$, whereas for d =\SI{300}{nm}, the synchronized frequency is higher than $f_{2}$. Additionally, synchronization is stronger at shorter separations, leading to a frequency shift of \SI{0.5}{GHz} and \SI{0.2}{GHz} for \SI{100}{nm} and \SI{300}{nm} separation, respectively. 
Synchronization in these devices is mediated by spin wave propagation, governed by two key factors. First, the phase difference between the two waves can lead to either in-phase or anti-phase synchronization, manifesting as constructive or destructive interference. Second, the attenuation of the spin-wave amplitude, which decays as ($A\exp(-r/L_{att})$), with A is the amplitude and $L_{att}$ is the attenuation length, limits synchronization over longer distances, with a cutoff around 600 nm in the present devices. Figs. 3(e) and 3(f) display the spatial intensity profile of the FFT of $m_{z}$, revealing the spin wave interference pattern in the synchronized state. For d =\SI{100}{nm}, synchronization occurs in phase, as indicated by a maximum interference pattern at the center between the two oscillators. In contrast, for d =\SI{300}{nm}, synchronization occurs in an anti-phase configuration, with a minimum at the center. These findings demonstrate that the separation distance dictates whether SHO1 and SHO2 synchronize in phase or anti-phase.
\begin{figure*}[t]
\begin{center}
\includegraphics[width=0.7\textwidth]{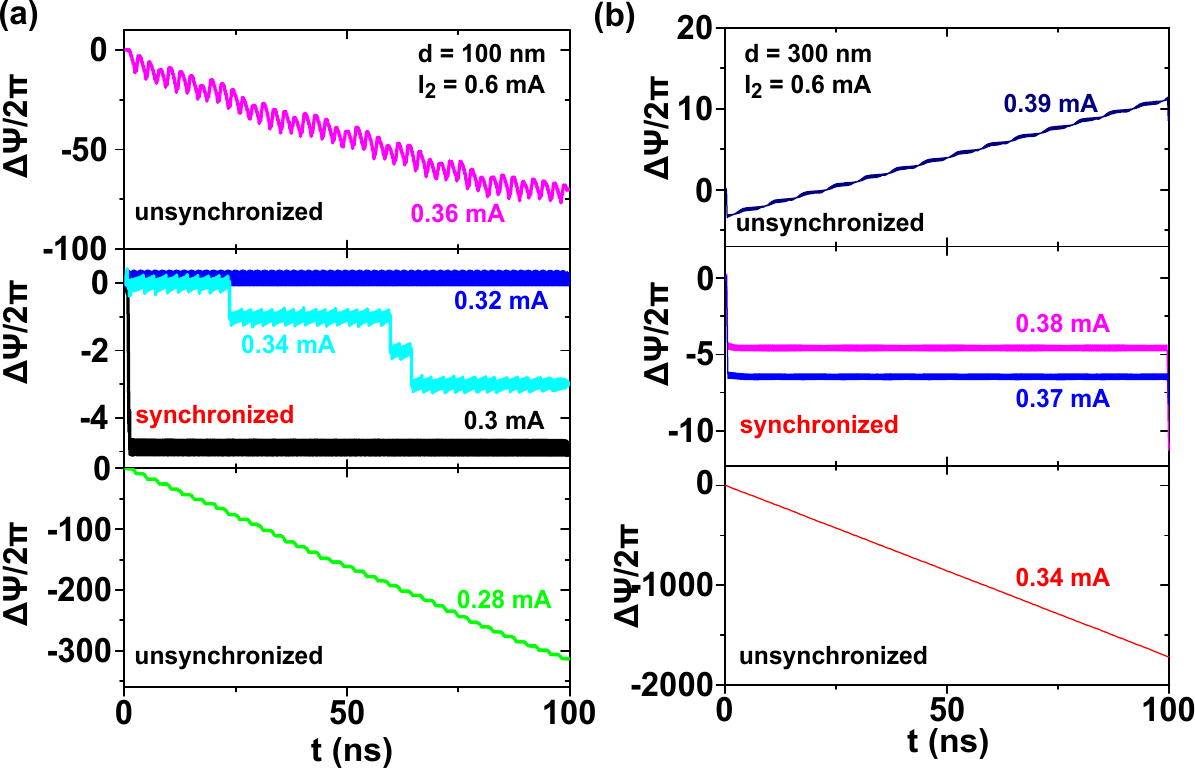}
\caption{(Color online) The instantaneous phase difference ($\Delta \Psi$) between SHO1 and SHO2 as a function of time for different values of the direct current $I_{1}$ applied to SHO1, for separation distances of (a) \SI{100}{nm} and (b) \SI{300}{nm}. The direct current applied to SHO2 is fixed at $I_{2}=$ \SI{0.6}{mA}. }
\end{center}
\end{figure*}
To gain deeper insight into the synchronization process between the two oscillators, we compute the phase of each oscillator individually. Using the time traces of $m_{z}$, we determine the instantaneous phase difference between the oscillators following the procedure described in \cite{Bianchini2010}:
\begin{equation}
    v_{a}(t) = v(t) + j H[v{t}] = A(t) e^{\Psi(t)} 
\end{equation}

where $j$ is the imaginary unit, $H[v(t)]$ represents the Hilbert transform of the time trace, $A(t)$ is the instantaneous amplitude, and $\Psi (t)$ is the instantaneous phase of $v(t)$. We independently record the dynamics of each oscillator and extract $\Psi_{1} (t)$ for SHO1 and $\Psi_{2} (t)$ for SHO2. Figs. 4(a) and 4(b) present the phase difference $\Delta \Psi (t) = \Psi_{1} (t) - \Psi_{2} (t)$ calculated from 100 ns time traces for different values of $I_{1}$ in devices with d =\SI{100}{nm} and \SI{300}{nm}. In theory, phase-locked synchronization requires that the phase difference remains constant over time, i.e., $\Delta \Psi (t) = \Psi_{1}- \Psi_{2} =$ constant \cite{Slavin2006}. At low currents of $I_{1}$ (bottom panel), the instantaneous phase difference varies over time with a negative slope, indicating a fully unsynchronized state. As $I_{1}$ increases (central panel), $\Delta \Psi$ stabilizes, signifying a transition from an unsynchronized state to a phase-locked synchronized state. Notably, at transition currents (e.g., for $I_{1}= $ \SI{0.34}{mA}), $\Delta \Psi$ exhibits discrete steps over time. Although both oscillators oscillate at the same frequency, their mutual coupling is insufficient to maintain fully locked phase synchronization, leading to periodic desynchronization-resynchronization events over short timescales of approximately \SI{30}{ns}. Additionally, we observe that synchronization occurs at a random phase difference $\Delta \Psi$ rather than the theoretically expected values of 0 or $\pi$. This deviation arises because SHO1 and SHO2 are not identical, leading to an inherent phase offset in their synchronization. At higher current (top panel), the two oscillators oscillate with significantly different phases, leading to an unsynchronized state. This device has several advantages: (i) The local damping is high in the oscillator region, where the propagating wave is initiated, and low outside in the ferromagnet, resulting in larger propagation distances compared to the common spin Hall nano-oscillator layout such as nano-gap or nano-constriction layout. (ii) This layout provides an independent measurement of the phase of the oscillators using, for example, an oscilloscope. Additionally, the fact that two currents can pass through the device offers an extra control parameter for tuning the synchronization process between the two oscillators.

In conclusion, we investigated the synchronization of propagating spin waves in a novel spin-torque oscillator (STO) layout using micromagnetic simulations. This device configuration enables independent probing of the dc current in each oscillator, allowing for precise control of the synchronization state and direct measurement of the oscillator phase. We analyzed the power spectral density of individual oscillators and identified the conditions under which synchronization occurs. By systematically varying the current in one oscillator while keeping the other fixed, we observed a transition from an unsynchronized state to a phase-locked state. The frequency locking was accompanied by characteristic interference patterns in the spin-wave intensity, confirming the role of propagating spin waves in mediating synchronization. Furthermore, phase analysis revealed that the instantaneous phase difference remains constant in the synchronized state. 
Our findings provide valuable insights into the synchronization mechanisms of spin Hall nano-oscillator and highlight the influence of separation distance on phase relationships. This work paves the way for further experimental validation and the development of spin-torque based technologies for neuromorphic computing, frequency generation, and energy-efficient spintronic devices.

\begin{acknowledgments}
This work is supported by the American University of Beirut Research Board (URB).
\end{acknowledgments}

\textbf{DATA AVAILABILITY}

The data that support the findings of this study are available from the corresponding author upon reasonable request.

%

\end{document}